\documentclass[]{interact}

\usepackage{epstopdf}
\usepackage[caption=false]{subfig}

\usepackage[numbers,sort&compress,merge]{natbib}
\bibpunct[, ]{[}{]}{,}{n}{,}{,}
\renewcommand\bibfont{\fontsize{10}{12}\selectfont}

\theoremstyle{plain}

\theoremstyle{definition}

\theoremstyle{remark}

\begin{document}

\articletype{ARTICLE}

\title{Simulating Passage through a Cascade of Conical Intersections with Collapse-to-a-Block Molecular Dynamics}

\author{
\name{Fangchun Liang\textsuperscript{a} and Benjamin G. Levine\textsuperscript{a}\thanks{CONTACT Benjamin G. Levine. Email: ben.levine@stonybrook.edu}}
\affil{\textsuperscript{a}Department of Chemistry and Institute for Advanced Computational Science, Stony Brook University, Stony Brook, NY  11794}
}

\maketitle

\begin{abstract}
The Ehrenfest with collapse-to-a-block (TAB) molecular dynamics approach was recently introduced to allow accurate simulation of nonadiabatic dynamics on many electronic states.  Previous benchmarking work has demonstrated it to be highly accurate for modeling dynamics in one-dimensional analytical models, but nonadiabatic dynamics often involves conical intersections, which are inherently two-dimensional.  In this report, we assess the performance of TAB on two-dimensional models of cascades of conical intersections in dense manifolds of states. Several variants of TAB are considered, including TAB-w, which is based on the assumption of a Gaussian rather than exponential decay of the coherence, and TAB-DMS, which incorporates an efficient collapse procedure based on approximate eigenstates.  Upon comparison to numerically exact quantum dynamics simulations, it is found that all TAB variants provide a suitable description of the dynamical passage through a cascade of conical intersections.   The TAB-w approach is found to provide a somewhat more accurate description of population dynamics than the original TAB method, with final absolute population errors  $\leq$0.013 in all cases.  Even when only four approximate eigenstates are computed, the use of approximate eigenstates was found to introduce minimal additional error (absolute population error $\leq$0.018 in all models).
\end{abstract}


\section{\label{sec:level1}Introduction}
Conical intersections, points of degeneracy between two or more adiabatic electronic states, are now widely known to be important objects for describing and understanding photochemistry, akin to transition states in ground state chemistry \cite{yarkony1996diabolical,levine2007isomerization,zhu2016review,schuurman2018dynamics, levine2019conical,matsika2021chemrev}.  
When a molecule passes through or near a conical intersection, the probability of transitioning between states is near unity.  
Conical intersections exist as $(N-2)$-dimensional seams in nuclear configuration space, where $N$ is the total number of nuclear degrees of freedom.  The dynamics of a molecule passing near a conical intersection are heavily influenced by motion in the two remaining degrees of freedom.\cite{yarkony2001dynamics,cederbaum2005shorttime}.
These degrees of freedom, known together as the branching space, include the nonadiabatic coupling direction and the energy difference gradient direction.  
Nuclear motion in the nonadiabatic coupling direction couples populations on the upper and lower PESs, resulting in transitions between adiabatic electronic states. 
The difference gradient direction drives populations on different states in different directions, resulting in the bifurcation of the population.  
Both branching directions vary rapidly with position near the intersection, adding to the difficulty of solving for the dynamics.     

Exact quantum dynamical simulation of nonadiabatic dynamics around a conical intersection is only feasible in systems with relatively few degrees of freedom. Therefore, in large polyatomic molecules, approximations that mix quantum and classical dynamics are often used.  
The need to accurately describe bifurcation of the population at the intersection places demands on these approximations.  
Bifurcation of the population is well captured by methods such as trajectory surface hopping \cite{tully1971trajectory,tully1990molecular, hammes1994proton,coker1995methods,tully1998mixed,jasper2001photodissociation, subotnik2016understanding,wang2016recent}, Gaussian wave packet methods such as spawning \cite{ben2000ab,worth2003full, wu2003matching,curchod2018ab}, path integral methods \cite{makri2015quantum,walters2015quantum}, exact factorization-derived methods \cite{min2015coupled,min2017ab,curchod2018ct,gossel2018coupled,vindel2022exact} as well as other fully quantum approaches \cite{kouppel1984multimode,guo2016accurate}.
Ehrenfest, or mean-field, molecular dynamics  \cite{miller1978classical,meyera1979classical,macias1982ab,micha1983self,cotton2013symmetrical,isborn2007time} is well known to fail to describe bifurcation of the wavepacket between ground and excited state following passage through an intersection \cite{tully1990molecular}.  
As such, several decoherence corrections to Ehrenfest dynamics have been developed to reproduce the bifurcation of the exact wave function.  These decoherence corrections take one of several forms: a) a stochastic collapse of the electronic wave function into a particular adiabatic electronic state\cite{prezhdo1997mean, prezhdo1997evaluation, bedard2005mean, subotnik2010augmented}, b) a gradual drift of the electronic wave function towards a particular adiabatic electronic state\cite{hack2001natural, zhu2004non, zhu2004coherent, akimov2014coherence}, or c) a fully quantum treatment based on coupled frozen Gaussian basis functions that evolve according to Ehrenfest equations of motion\cite{makhov2014ab, freixas2018ab, fedorov2019nonadiabatic}.  In all cases, the method is designed such that the distribution of outcomes over an ensemble of trajectories resembles that which would be derived from an exact quantum dynamical calculation.

The above-described methods are generally well-suited for simulating systems that occupy a small number of intersecting electronic states.  However, there are many interesting problems, from molecular physics \cite{kaufman2023coherence,holland2024auger} to materials science \cite{alzubeidi2023solvatede,hetherington2023cooling}, that involve dynamics in dense manifolds of intersecting electronic states.  Describing nonadiabatic dynamics in such systems poses a unique challenge that has drawn the attention of many researchers \cite{craig2005trajectory,dou2015surface,dou2015frictional,ouyang2015surface,fedorov2019nonadiabatic,esch2020decoherence}.  On the one hand, methods that simplify the treatment of dense manifolds of states by integrating over a bath of states or using a single-particle representation of the electronic structure can be extremely valuable tools for specific applications.  However, the detailed dynamics around conical intersections are lost in both cases.  On the other hand, traditional few-state methods such as trajectory surface hopping, spawning, and decoherence-corrected Ehrenfest methods typically require knowledge of the full electronic eigenspectrum at all time steps, which is impractical for systems with very dense manifolds of states.

With the challenge of simulating dynamics in dense manifolds of intersecting states in mind, our group has recently developed the Ehrenfest with collapse-to-a-block (TAB) family of methods.  TAB is based on a state-pairwise description of decoherence, which can discriminate between the long-lived coherences of nearly parallel electronic states and the short-lived coherences of non-parallel pairs of states  \cite{esch2020state}.  As described below, the extension of TAB to incorporate an inexpensively obtained approximate eigenstate basis \cite{fedorov2019nonadiabatic} enables the accurate description of bifurcation near surface crossings without fully diagonalizing the Hamiltonian at each time step. \cite{esch2020decoherence}  In the context of a recent community challenge to predict the ultrafast electron diffraction spectrum of cyclobutanone, we have employed an \textit{ab initio} molecular dynamics implementation of TAB for the first time \cite{suchan2024predict}.

In the past, we have benchmarked the accuracy of various flavors of TAB relative to exact quantum dynamical simulations by application to several one-dimensional model potentials \cite{esch2020state,esch2020decoherence,esch2021accurate}.  However, conical intersections are inherently two-dimensional objects.  Thus, in this work we investigate the accuracy of several TAB methods for describing the dynamics for several two-dimensional models of cascades of conical intersections.  In section \ref{sec:methods} we describe the family of TAB methods, the specific analytical model Hamiltonians we will use in our study, and other computational details.  In section \ref{sec:results}, we analyze the accuracy of our TAB simulations compared to exact quantum dynamical simulations.  Finally, in section \ref{sec:conclusions}, we draw conclusions and discuss future prospects.

\section{Methods} \label{sec:methods}
Here we present an overview of the family of TAB methods, followed by a description of the model Hamiltonians used in this work and the computational details of the simulations.

\subsection{TAB Family of Methods} \label{ss:family}

The TAB methods each comprise two familiar methodological components: an Ehrenfest propagation scheme, and a correction for decoherence, which is applied at the end of each classical time step.  The decoherence correction takes the form of a stochastic collapse, akin to the mean-field with stochastic decoherence method \cite{bedard2005mean}. 
The distinct feature of TAB compared to other related methods is that it allows collapse to a superposition of states.
In so doing, coherences between some pairs of states may be long lived, while others may be lost quickly, just as would occur during exact quantum propagation.
Several variants exist, differing in the specifics of the decoherence correction.
The performance of several variants will be analyzed in this work, thus, here we detail the family of TAB algorithms, emphasizing the universal features as well as the differences.  The key differences are summarized in Table \ref{tbl:family} and discussed in detail below.

\begin{table}
  \caption{Differentiating features of the TAB family of methods.}
  \label{tbl:family}
  \begin{tabular}{ccc}
    \hline
      Method & Eigenstate Basis & Coherence Decay \\
    \hline
    TAB        & Exact       & Exponential \\
    TAB-DMS    & Approximate & Exponential \\
    TAB-w      & Exact       & Gaussian    \\
    TAB-w-DMS  & Approximate & Gaussian    \\
    \hline
  \end{tabular}
\end{table}

First we briefly outline the key aspects of the Ehrenfest propagation, which is identical to that used on our past work \cite{esch2020decoherence}. 
We use the velocity Verlet algorithm \cite{swope1982computer} for nuclear motion with a time step $\Delta t$ selected by the user. 
A second-order symplectic split operator propagator \cite{blanes2006symplectic}, with a shorter electronic time step $\Delta t_e$, is employed to propagate electronic amplitudes $\textbf{c}(t)$. 

At the end of each classical time step, the decoherence correction is applied in a series of two steps: a) a target mixed state density matrix is generated, and b) the wave function stochastically collapses into a coherent superposition of electronic states, such that the expectation value of the resulting pure state density matrices is equal to the target density matrix.  
These two steps will be described in subsections \ref{sss:target} and \ref{sss:collapse}.  
But first, in subsection \ref{sss:basis}, we will discuss the choice of basis in which to generate the target density matrix.

\subsubsection{Choice of Basis}\label{sss:basis}

The TAB family of methods has been conceived to be used in an \textit{ab initio} molecular dynamics context, performing electronic structure calculations on the fly.  
In the limit where a relatively small number of electronic states are populated, it is therefore most natural to apply TAB in the adiabatic basis.  Our earliest work on TAB used the adiabatic basis, and where not specified otherwise, this is the default.

However, the ultimate goal of developing the TAB family of methods is to be able to perform accurate nonadiabatic molecular dynamics simulations where the number of populated states is large.  
In this limit, diagonalization of the Hamiltonian at each time step is cumbersome.  
A key advantage of Ehrenfest propagation is that it does not require such diagonalization on its own.  
In order to apply TAB in this limit, we have developed a procedure to compute an approximate adiabatic basis from the history of the mean-field wave function.  When approximate eigenstates are used for the collapse, we call our method TAB for dense manifolds of states, or TAB-DMS.

The approximate adiabatic basis is determined at the end of each classical time step.  
To this end, the mean-field electronic wave function is stored at several times within the classical time step (in the present case, separated by intervals of 0.0040 a.u.).  
Each complex wave function is separated into two real vectors at each time, $\boldsymbol{c}_R(t)$ and $\boldsymbol{c}_I(t)$, corresponding to the real and imaginary parts of the mean field wave function, according to
\begin{align}
    \boldsymbol{c}(t)=\boldsymbol{c}_R(t)+i\boldsymbol{c}_I(t).
\end{align}
This basis is orthogonalized, and then the Hamiltonian is built and diagonalized in the resulting basis to generate a set of approximate eigenstates that will be used to construct the target density matrix.  
This basis is akin to the Krylov subspace of the Hamiltonian acting on the current mean-field wave function, and thus is a reasonable basis in which to search for approximate eigenstates. 
Importantly, the current electronic wave function is guaranteed to be exactly representable as a linear combination of the approximate eigenstates.
In practical tests in one-dimensional models, we have found that only a small number of approximations states (often only $\sim$4) are  needed to converge TAB-DMS simulations to the exact TAB result \cite{esch2020decoherence}.
TAB-DMS reduces to full TAB in the case that the number of approximate eigenstates is the full dimension of the electronic Hilbert space.

\subsubsection{Generation of Target Density Matrix}\label{sss:target}

Having chosen a basis, at each time step we will generate the target density matrix in that basis. 
The target density matrix, $\boldsymbol{\rho}^d$, will describe a mixed state that incorporates the decoherence between states in a state-pairwise fashion.  
We start from the current coherent density matrix, $\boldsymbol{\rho}^c$, represented in either the adiabatic or approximate adiabatic basis.
The diagonal elements of the target density matrix must match this coherent density matrix so that the expectation values of the state populations are conserved during collapse,
\begin{align}
    \rho_{ii}^d = \rho_{ii}^c(t_0+\Delta t).
\end{align}
The off diagonal elements, however, will be scaled by a real, positive scalar between zero and one to reflect the effects of decoherence.

In this work we will use two different definitions of this scaling factor.  
In our original formulations of TAB and TAB-DMS, we assume that the coherences decay exponentially,
\begin{align}
    \rho_{ij}^d = \rho_{ij}^c(t_0+\Delta t)e^{-\frac{\Delta t}{\tau_{ij}}}.
\end{align}
The state-pairwise decoherence times, $\tau_{ij}$, are computed according to the classic formula derived by Bittner and Rossky \cite{bittner1995decoherence}.
\begin{align}
    \tau_{ij} = \sum\frac{(F_{i,\eta}^{avg}- F_{j,\eta}^{avg})^2}{8\hbar^2\alpha_{\eta}}.
\end{align}
Here $\eta$ indexes the nuclear degrees of freedom, $F_{i,\eta}^{avg}$ is an element of the force vector of state $i$, and $\alpha_{\eta}$ is a decoherence parameter (in units of inverse length squared) that corresponds to the width of the vibrational wave packet.

The Bittner-Rossky decoherence time was derived originally as the width of a Gaussian decay, arising from the overlap of a pair of Gaussian wave packets on non-parallel PESs.  Thus, a second natural option is to assume Gaussian decay of the coherences.  The challenge of incorporating Gaussian decay is that, unlike an exponential, the rate of Gaussian decay depends on the history of the wave function.  Where the derivative of an exponential decay is proportional only to the current value of the function, 
\begin{align}
    \frac{d}{dt}e^{-\frac{t}{\tau_{ij}}}=-\tau_{ij}^{-1}e^{-\frac{t}{\tau_{ij}}},
\end{align}
for Gaussian decay, the relative rate of decay is initially zero, and increased with increasing time,
\begin{align}
    \frac{d}{dt}e^{-\frac{t^2}{\tau_{ij}^2}}=\frac{-2t}{\tau_{ij}^{2}}e^{-\frac{t^2}{\tau_{ij}^2}}.
\end{align}
The result is that using a Gaussian decay requires knowledge of the history of the coherence.  That is, one must know how long ago the coherence was created to know its rate of decay.  To this end, we have proposed and tested a scheme to approximate that history that requires only information from the current time step \cite{esch2021accurate}.  This approximation was referred to as TAB-w1 in that previous work.  Herein we simply refer to the resulting methods as TAB-w and TAB-w-DMS, when Gaussian decay is employed along with exact and approximation eigenstates, respectively.  The reader is referred to reference \cite{esch2021accurate} for the detailed algorithm.

\subsubsection{Wave Function Collapse}\label{sss:collapse}

Once that target density matrix has been generated, we will stochastically collapse the wave function into a superposition of one or more adiabatic electronic states.  
To this end, the target density matrix is expanded as a sum of pure-state density matrices representing such superposition states,
\begin{align}
    \boldsymbol{\rho}^d = \sum_a P_a^{block} \boldsymbol{\rho}_a^{block}.
\end{align}
Here $a$ indexes the set of possible superpositions, and $P_a^{block}$ is the probability that the mean-field electronic state will collapse into the pure state described by $\boldsymbol{\rho}_a^{block}$ in the present time step.  The algorithm for determining $\{\boldsymbol{\rho}_a^{block}\}$ and $\{P_a^{block}\}$ is presented in detail in Ref \cite{esch2021accurate}.
The specific procedure we use here was referred to as rTAB in this past work, which compares several algorithms for expanding $\boldsymbol{\rho}^d$.

Though we refer the reader to this past work for implementation details, here we note several important facts about the expansion of $\boldsymbol{\rho}^d$.  
Each $\boldsymbol{\rho}_a^{block}$ represents a pure superposition of a subset of the basis electronic states (adiabatic or approximately adiabatic).
In each superposition, the total population must be unity and the relative phases of the individual basis states are the same as in $\boldsymbol{\rho}^c$.  
So, for example, consider a system with three electronic states.  The pure mean-field state may initially be represented as,
\begin{align}
    \boldsymbol{\rho}^c = \left[ {\begin{array}{c}
   c_0 \\
   c_1 \\
   c_2 \\
  \end{array} } \right]
  \left[
  {\begin{array}{ccc}
   c_0^* & c_1^* & c_2^* \\
  \end{array} } \right].
\end{align}
Possible superposition states include the state in which all three states remain populated,
\begin{align}
    \boldsymbol{\rho}_a^{block} = \boldsymbol{\rho}^c = \left[ {\begin{array}{c}
   c_0 \\
   c_1 \\
   c_2 \\
  \end{array} } \right]
  \left[
  {\begin{array}{ccc}
   c_0^* & c_1^* & c_2^* \\
  \end{array} } \right],
\end{align}
states where only two states remain populated, for example,
\begin{align}
    \boldsymbol{\rho}_a^{block} =
    \frac{1}{|c_0|^2+|c_1|^2}
    \left[ {\begin{array}{c}
   c_0 \\
   c_1 \\
   0 \\
  \end{array} } \right]
  \left[
  {\begin{array}{ccc}
   c_0^* & c_1^* & 0 \\
  \end{array} } \right],
\end{align}
or states in which only a single state remains populated, for example,
\begin{align}
    \boldsymbol{\rho}_a^{block} = \left[ {\begin{array}{c}
   0 \\
   0 \\
   1 \\
  \end{array} } \right]
  \left[
  {\begin{array}{ccc}
   0 & 0 & 1 \\
  \end{array} } \right].
\end{align}
In the above, $\bold{c}$ represents the mean-field wave function expansion coefficients prior to collapse.  Though there are $2^M-1$ possible elements of $\{P_a^{block}\}$, where $M$ is the number of basis states, in practice we employ a greedy algorithm that generates an expansion of $M(M+1)$ blocks in $\mathcal{O}(M^2)$ time, avoiding exponential scaling.  This expansion is uniquely defined, but in very rare instances it does not exactly reproduce $\boldsymbol{\rho}^d$.

To collapse, a random number between zero and one is chosen and compared to $\{P_a^{block}\}$ to choose a random block.  The mean-field wave function is then reassigned to reflect the collapse.
The wave function collapse changes the electronic energy of the system. 
Similar to other nonadiabatic molecular dynamics methods, we rescale the momentum to ensure energy conservation.
In the present work, we rescale the entire momentum vector such that total energy is conserved.  In future work, we will investigate more sophisticated approaches for momentum rescaling, more akin to those that are preferred for use with trajectory surface hopping \cite{subotnik2011rescaling,martens2019rescaling,plasser2019rescaling,barbatti2021rescaling}.  
The energetics of our models are such that we need not be concerned with energetically forbidden (``frustrated'') collapses in this work, though this will also be a direction of future study.

\subsection{Model Hamiltonians and Simulation Details}
In order to assess their accuracy, we have applied ordinary Ehrenfest and the variants of TAB to a series of two-dimensional analytical model systems.  All simulations are compared to numerically exact quantum dynamical calculations as a point of reference. 
In contrast to our recent application of an \textit{ab initio} implementation of TAB to predict an experimental ultrafast spectrum \cite{suchan2024predict}, comparison to exact quantum dynamical simulations on model potentials enables us to isolate errors associated with the dynamical approach and avoid the many challenges associated with accurately computing the PES near conical intersections \cite{nangia2005balanced,levine2006tddft,granovsky2011xms,gozem2014shape,park2021butadiene}.

\begin{figure*}
    \centering
    \includegraphics[width=1\linewidth]{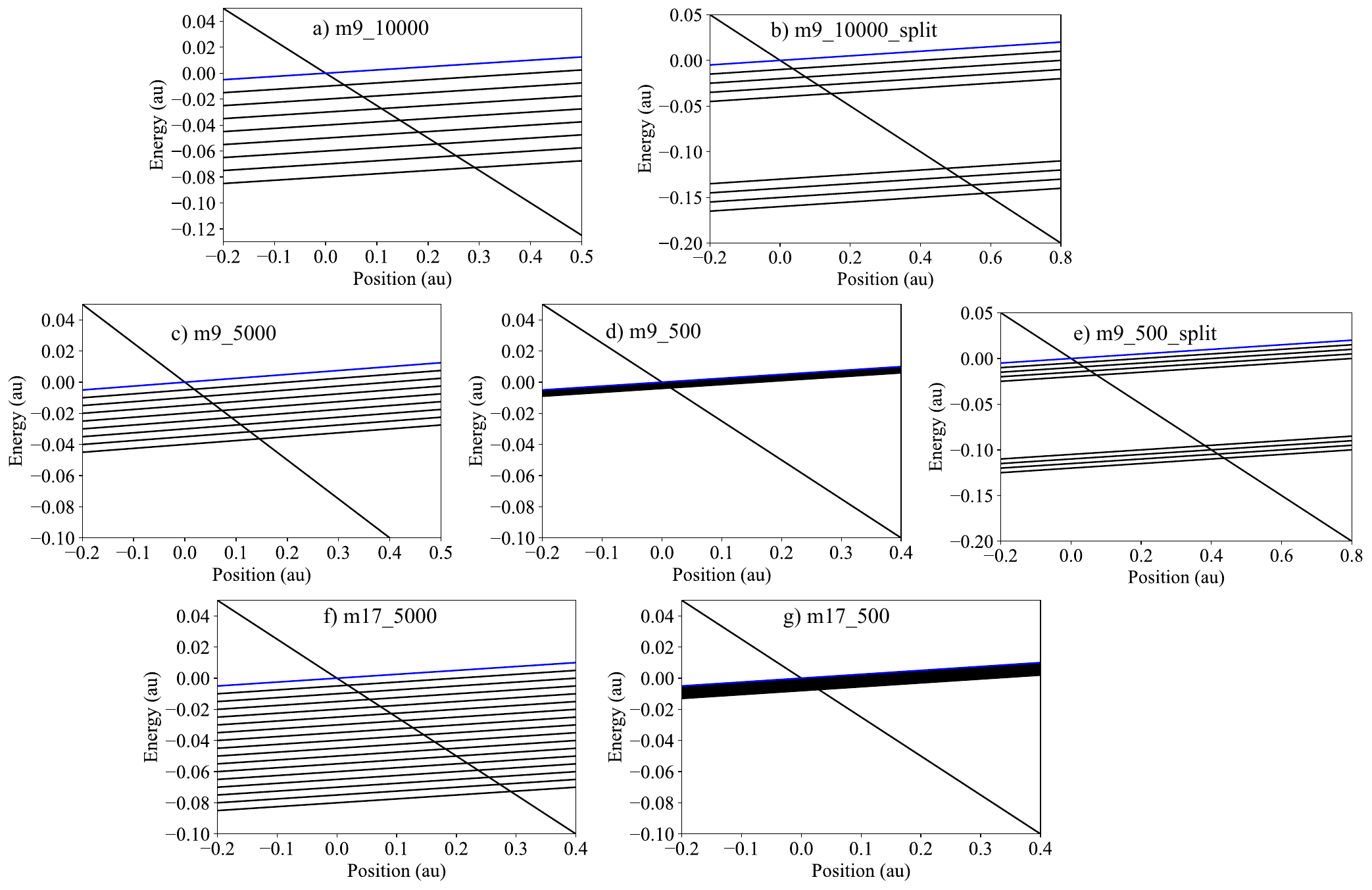}
    \caption{One-dimensional slices of the diabatic PESs of the seven two-dimensional models (for $y=0$).}
    \label{fig:PES}
\end{figure*}

Each model represents passage through a series of conical intersections that arise from the crossing of a sloped diabatic electronic state (which we hereafter refer to as the first diabatic state) with a band of parallel diabatic states.  
The series of models includes three nine-state models, two seventeen-state models, and another two nine-state models in which a large gap separates two bands of parallel states.  One-dimensional slices of these models are depicted in Figure \ref{fig:PES}, where the black lines indicate the diabatic state energies on the $x$-direction and the blue lines indicate the diabatic coupling on the $y$-direction.

The electronic Hamiltonian is defined in the diabatic basis. The first diabatic state has negative slope in the $x-$direction,
\begin{align}
  H_{11} = -w_1 x,
\end{align}
where $w_1$ has a value of 0.25 a.u. for all models.
The other states (2 to $M$) are parallel to one another, forming a band of dense states. 
For the nine- and seventeen-state models without a split band, the diagonal matrix elements corresponding to the energies of these states are 
\begin{align} \label{eq:Hii}
  H_{ii} = w_2 x -(i-1)\delta,\, i\in[2,M],
\end{align}
where the slope, $w_2$, has a value of 0.025 a.u. for all models.  The energy spacing between states is $\delta$.
For the nine-state split models, states 2-5 are as in Eq. \ref{eq:Hii}.  States 6-9 remain parallel, but are separated by a large gap, with diagonal matrix elements:
\begin{align}
  H_{ii} = w_2 x -(i-1)\delta-\epsilon, \,i\in[6,9].
\end{align}
The first diabatic state is coupled to the other states with a coupling defined
\begin{align}
  H_{1i}=H_{i1} &=ky,\, i\in[2,M].
\end{align} 
The coupling constant is $k$= 0.025 a.u. for all models.
The couplings between all pairs of parallel states are zero,
\begin{align}
  H_{ij}&=0,\,i\neq j,\, i\in[2,M],\,j\in[2,M].
\end{align}
The mass of the particle is 1822 a.u. in all models.
Table \ref{tbl:parameter} shows the $\epsilon$ and $\delta$ parameters for all seven models.
\begin{table}
  \caption{Model PES parameters}
  \label{tbl:parameter}
  \begin{tabular}{cccc}
    \hline
      & Number of States& $\delta$(microhartree) & $\varepsilon$ (microhartree)  \\
    \hline
    m9\_10000        & 9  & 10000 & 0 \\
    m9\_10000\_split & 9  & 10000 & 80000 \\
    m9\_5000         & 9  & 5000  & 0 \\
    m9\_500          & 9  & 500   & 0 \\
    m9\_500\_split   & 9  & 500   & 80000 \\
    m17\_5000        & 17 & 5000  & 0 \\
    m17\_500         & 17 & 500   & 0 \\
    \hline
  \end{tabular}
\end{table}

Numerically exact quantum mechanical simulations are implemented as described in earlier work \cite{fedorov2019nonadiabatic}. 
A two-dimensional space ranging from -4.0 a.u. to 8.0 a.u. in both the $x$- and $y$-directions are partitioned into (1000,1000) uniform grid points. 
For each model, a Gaussian nuclear wavepacket centered at (-1.0 a.u., 0.0 a.u.) with widths of 6.0 bohrs$^{-2}$ in both the $x$- and $y$-directions is initialized on the first diabatic state. 
The initial momentum was chosen to be 10.0 a.u. in both the $x$- and $y$-directions. 
Each single simulation was propagated for 400.0 a.u.

\begin{figure*}
    \centering
    \includegraphics[width=1\linewidth]{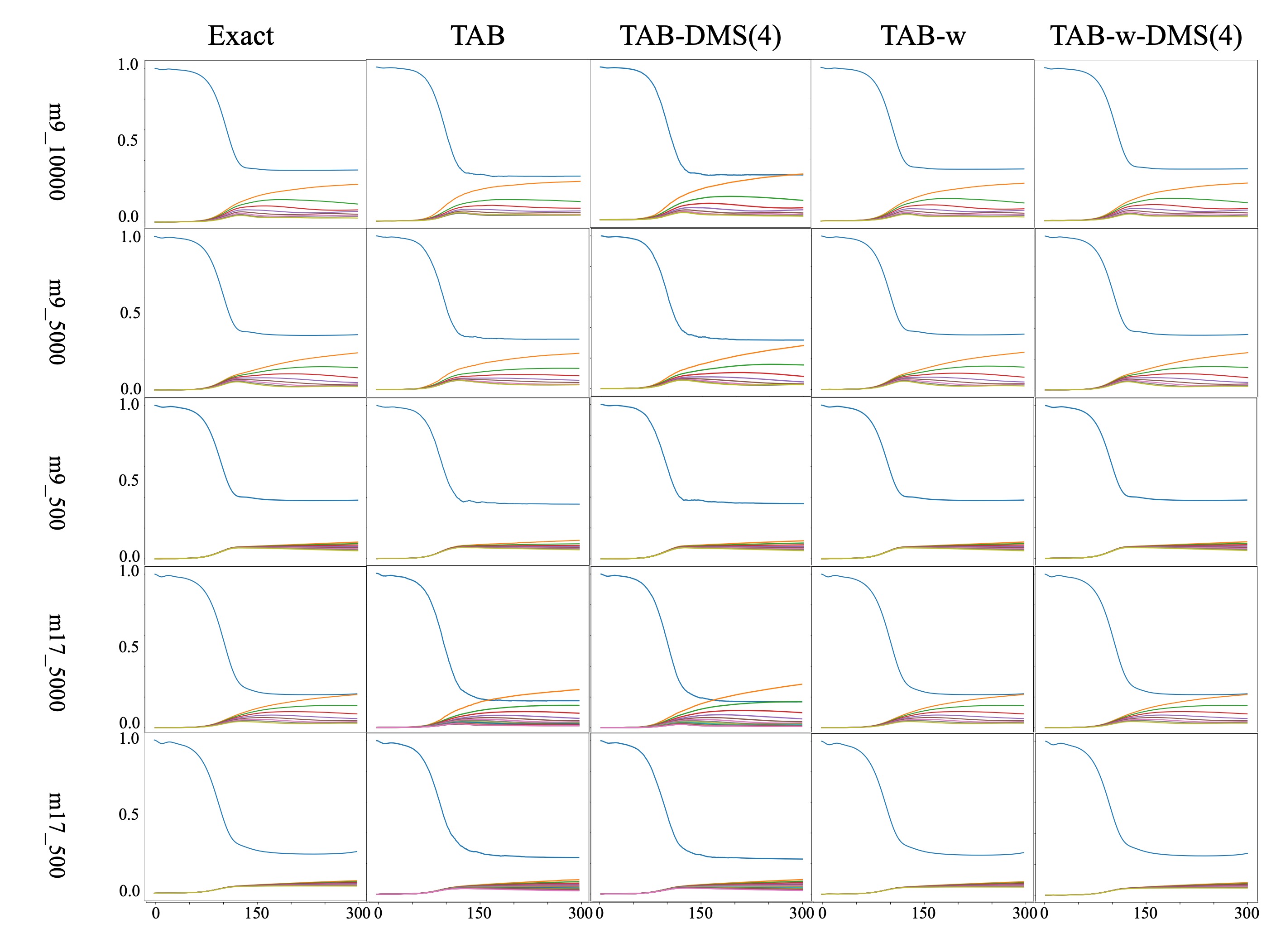}
    \caption{Time evolution of the diabatic state populations generated by exact quantum simulations compared to ensemble-averaged population evolution computed using TAB, TAB-DMS(4), TAB-w and TAB-w-DMS(4) methods for five even-spacing models.  Different colors correspond to different diabatic states.}
    \label{fig:pop1}
\end{figure*}

\begin{figure*}
    \centering
    \includegraphics[width=1\linewidth]{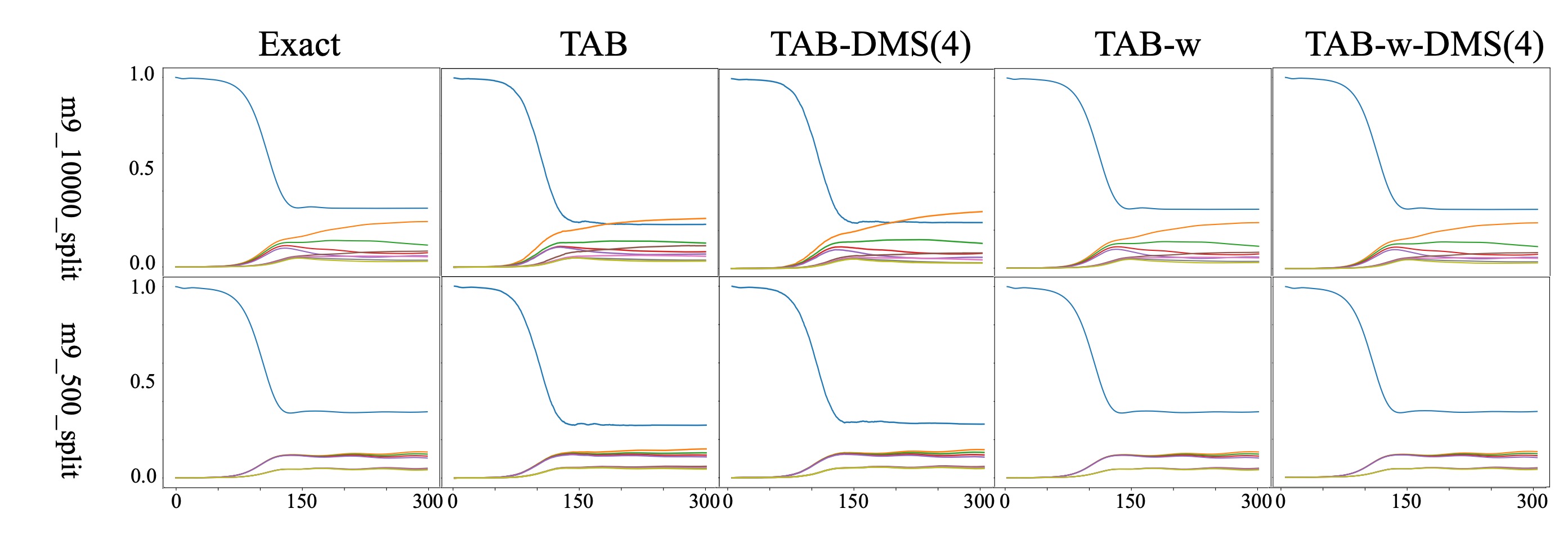}
    \caption{Time evolution of diabatic state populations generated by exact quantum simulations compared to ensemble-averaged population evolution using TAB, TAB-DMS(4), TAB-w and TAB-w-DMS(4) methods for the two 9-state split models.  Different colors correspond to different diabatic states.}
    \label{fig:pop2}
\end{figure*}

\begin{table*}
  \caption{Signed errors in the final first diabatic state population computed relative to numerically exact quantum results for all methods and models.  The smallest absolute error for each model is shown in bold.  TAB-DMS(4) indicates that four approximate eigenstates were used in this calculation.}
  \label{tbl:Pop}
  \begin{tabular}{cccccc}
    \hline
    Model            & TAB    & TAB-DMS(4) & TAB-w & TAB-w-DMS(4) & Ehrenfest\\
    \hline
    m9\_10000        & -0.046 & -0.044 & \textbf{-0.004} & -0.005 & -0.041\\
    m9\_10000\_split & -0.084 & -0.068 & \textbf{-0.013} & -0.018 & -0.047\\
    m9\_5000         & -0.033 & -0.043 &  \textbf{0.011} &  0.018 & -0.048\\
    m9\_500          & -0.026 & -0.025 & \textbf{-0.005} & -0.010 & -0.051\\
    m9\_500\_split   & -0.070 & -0.066 & \textbf{-0.006} & -0.009 & -0.054\\
    m17\_5000        & -0.048 & -0.053 &  \textbf{0.009} &  0.010 & -0.057\\
    m17\_500         & -0.033 & -0.044 & \textbf{-0.002} & -0.004 & -0.074\\
    \hline
  \end{tabular}
\end{table*}

For the standard Ehrenfest, TAB, TAB-DMS, TAB-w, and TAB-w-DMS simulations, the population evolution reported is averaged over 1000 individual trajectories. Initial nuclear positions and momenta of single trajectories were drawn from a Wigner distribution of the initial Gaussian nuclear wavepacket as in the numerically exact solver. Each trajectory is propagated for 400 a.u.  The classical and quantum degrees of freedom are propagated with time steps of $\Delta t=$ 0.05 a.u. and $\Delta t_e=$ 0.0005 a.u, respectively. The decoherence parameter in TAB was chosen to be $\alpha_x=\alpha_y=$ 6.0 bohrs$^{-2}$.  Note that this parameter is not empirically fitted. It was chosen to match the width of the initial Gaussian wave packets used in the exact quantum simulations.

\section{Results and Discussion} \label{sec:results}
Here we assess the performance of TAB from two perspectives. In subsection \ref{sub:population}, we will investigate the accuracy with which several TAB variants reproduce population dynamics, in comparison to numerically exact quantum simulations. In subsection \ref{sub:deviation}, we analyze how TAB rectifies the anomalous mean-field behavior observed in ordinary Ehrenfest dynamics following passage through the cascade of intersections.

\subsection{Population Dynamics}
\label{sub:population}
\begin{figure}
    \centering
    \includegraphics[width=1\linewidth]{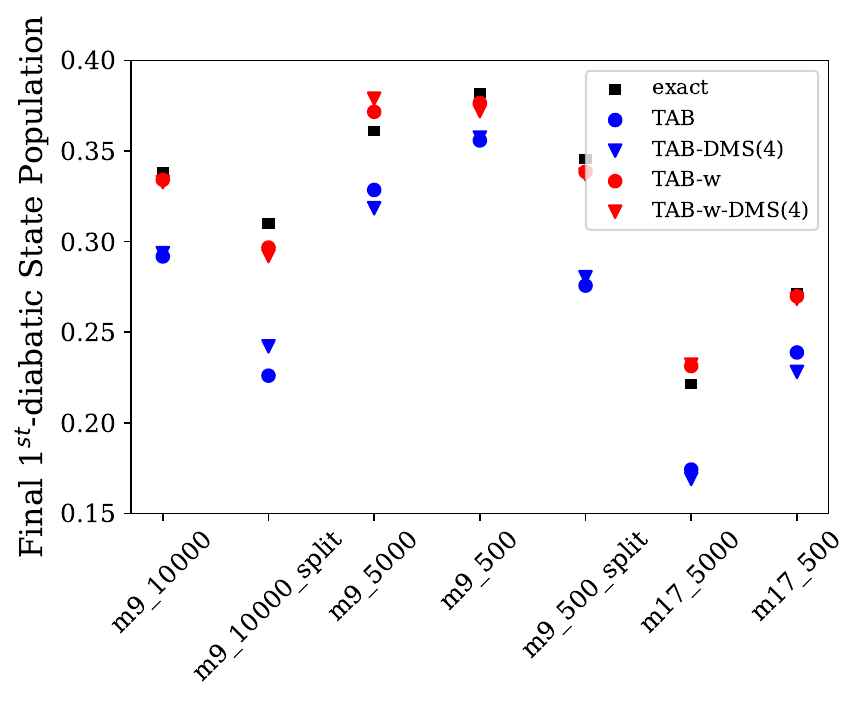}
    \caption{TAB (blue dots), TAB-DMS(4) (blue triangles), TAB-w (red dots), TAB-w-DMS(4) (red triangles), and numerically exact (black dashes) final first diabatic state populations for each of the seven model problems.}
    \label{fig:PopDev}
\end{figure}

Here, we are presenting the ensemble-average time-dependent diabatic state populations of TAB, TAB-DMS, TAB-w, and TAB-w-DMS for comparison to numerically exact quantum mechanical simulations.  The populations of all states are shown as a function of time in Figures \ref{fig:pop1} and \ref{fig:pop2}.  The final populations are summarized in Figure \ref{fig:PopDev}, which shows the averaged population of the first diabatic state at the end of the simulations.  In Table \ref{tbl:Pop}, the errors relative to exact quantum simulations are presented for all methods, including standard Ehrenfest. 

We can see that the TAB-w simulations have the smallest errors for all models.  The error observed in the final population of the first diabatic state is 0.013 in the worst case, and is below 0.010 for all but two models.  Using approximate eigenstates only slightly increases the error.  With only 4 approximate states, TAB-w-DMS provides the second most accurate results for all models, with all absolute errors below 0.020.  Figure \ref{fig:pop2} shows that the intricate population dynamics of the parallel diabatic states are also well described.

Assuming exponential decay of the decoherence results in larger absolute errors, ranging from 0.026 to 0.084 in the case of TAB.  Very similar errors are obtained with TAB-DMS when four approximate eigenstates are used, ranging from 0.025 to 0.068.  In all cases the signed errors are negative, indicating that TAB and TAB-DMS predict too much population transfer between diabats.  This suggests that the long tail of the exponential decay results in over-coherence, compared to the shorter tail of the Gaussian decay used in TAB-w.  For all four TAB methods, Figures \ref{fig:pop1} and \ref{fig:pop2} show that the populations closely track those of numerically exact quantum dynamical simulation for all states at all times.

The errors in the final population for uncorrected Ehrenfest simulations (shown only in table \ref{tbl:Pop}) are similar to those for TAB and TAB-DMS, ranging from 0.041 to 0.074.  Again, over-coherence results in too much population transfer between diabats. 

\begin{figure}
    \centering
    \includegraphics[width=1\linewidth]{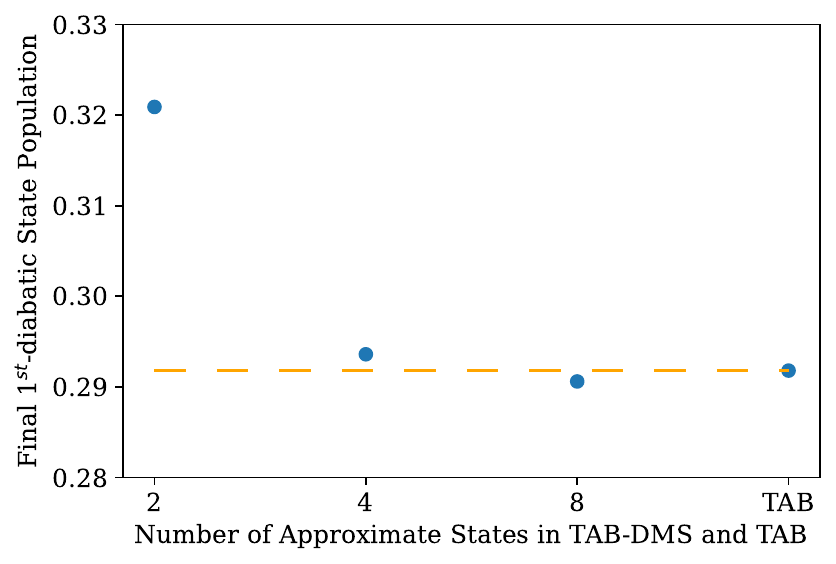}
    \caption{Final first diabatic state population for model m9\_10000 using TAB-DMS with 2, 4, 8 approximate adiabatic states as compared to TAB (with exact eigenstates).}
    \label{fig:conv}
\end{figure}

We also consider the convergence of TAB-DMS with respect to the number of approximate eigenstates. 
Figure \ref{fig:conv} shows the final population of the first diabatic state as a function of the number of approximate eigenstates. 
The TAB-DMS result has converged to the TAB result even when only 4 approximate eigenstates are used, with only small variations for larger numbers of approximate states.  This finding aligns with earlier published studies comparing TAB and TAB-DMS in one-dimensional models \cite{esch2020state, esch2021accurate} and multiple cloning for dense manifolds of states, which is based on the same approximate eigenstate basis \cite{fedorov2019nonadiabatic}.

\subsection{Deviation from Adiabatic Potentials}
\label{sub:deviation}
Ehrenfest dynamics can provide an accurate description of population dynamics near the intersection, but the pathological behavior of Ehrenfest is most visible after passing through the crossing region.  Once an Ehrenfest trajectory transitions into a superposition of multiple adiabatic states, there is no mechanism for coherence to be lost.  During this time, the trajectory evolves on an unphysical mean-field PES, and may fail to properly bifurcate between different reaction pathways.  As detailed in section \ref{ss:family}, the TAB method mitigates this deficiency by stochastically collapsing the electronic wave function. In order to quantify the extent to which TAB rectifies this flaw, we consider the deviation of the mean-field potential energy from the nearest adiabatic PES,

\begin{align}
  \Delta E = |E_p-E^a_{nearest}|.
\end{align}
Here $E_p$ is the mean-field potential energy and $E^a_{nearest}$ is the closest adiabatic potential at the current time step.

\begin{figure*}
    \centering
    \includegraphics[width=0.4\linewidth]{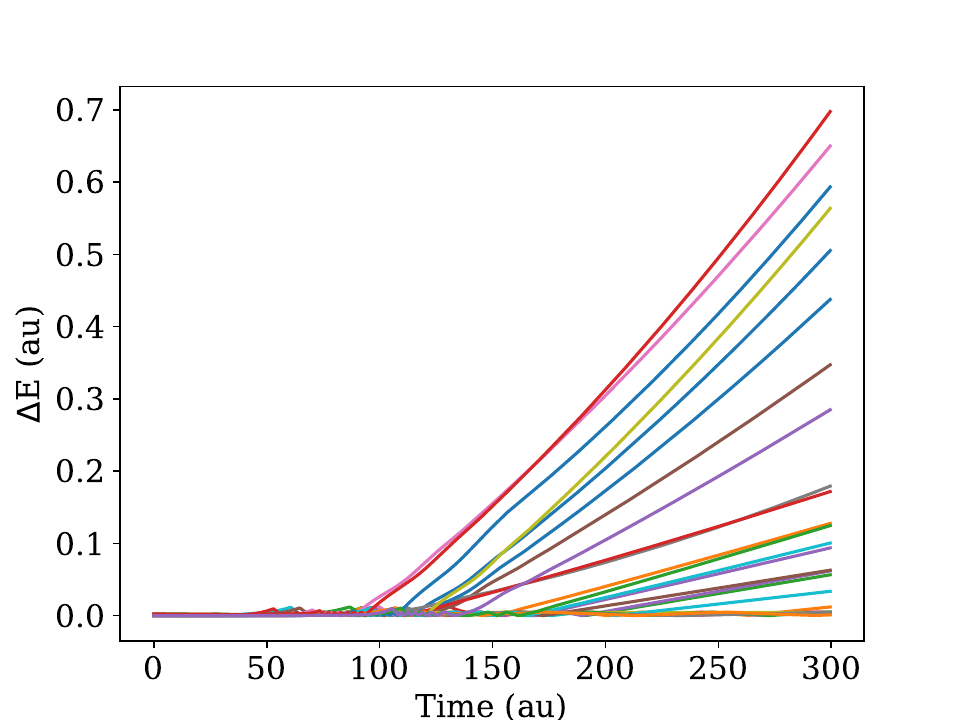}
    \includegraphics[width=0.4\linewidth]{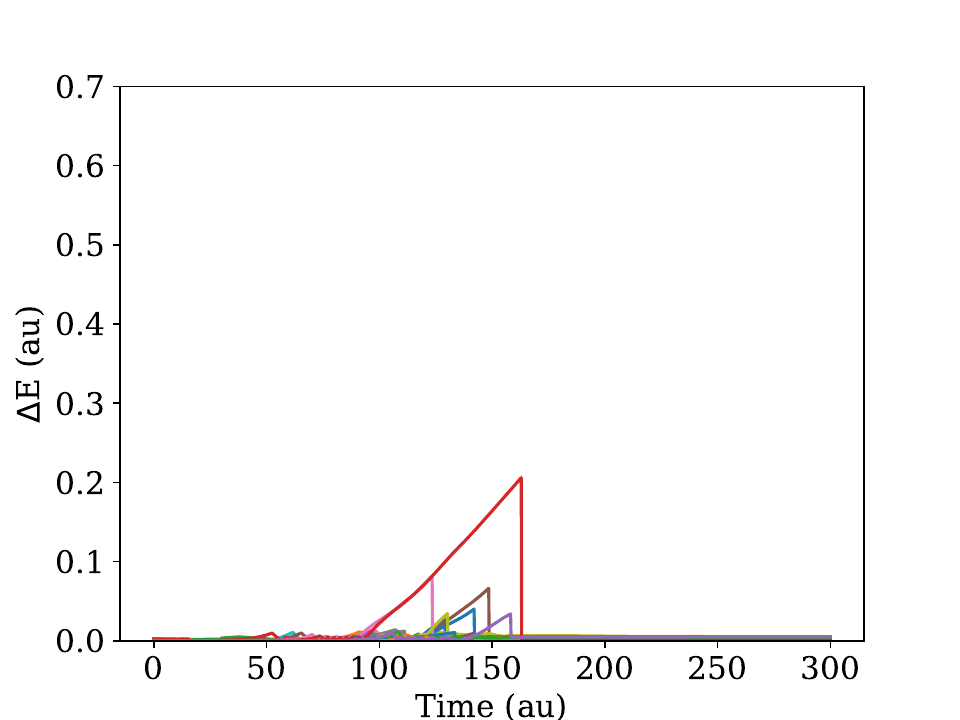}
    \caption{The deviation from the nearest adiabatic electronic PES, $\delta E$, as a function of time for twenty randomly chosen trajectories for model m9\_10000.  Results for ordinary Ehrenfest dynamics and TAB are presented on the left and right, respectively.}
    \label{fig:dev}
\end{figure*}

Figure \ref{fig:dev} presents $\Delta E$ as a function of time for several randomly selected Ehrenfest (left) and TAB (right) trajectories. In contrast to the conventional Ehrenfest dynamics, where nuclei tend to progressively diverge from the adiabatic PESs following passage through high-coupling regions, TAB electronic wavefunctions are effectively collapsed onto a specific adiabatic potential or a set of parallel adiabatic potentials. This stochastic collapse enables an ensemble of TAB trajectories to properly bifurcate, as an exactly propagated wavefunction would.

\section{Conclusions} \label{sec:conclusions}

In this work, we benchmarked the accuracy of several variants of the TAB method for simulating passage through a cascade of conical intersections between densely packed states.  The TAB approach is a variant of Ehrenfest molecular dynamics that incorporates a state-pairwise treatment of decoherence.  We find that all variants of TAB provide a suitable description of these dynamics.  The smallest population errors ($<0.02$ in all cases) were found for the TAB-w and TAB-w-DMS methods, which are based on a Gaussian (rather than exponential) decay of the coherence.  Somewhat larger population errors ($\sim0.05$) are found for the original TAB and TAB-DMS methods, which are based on an exponential decay.  The population errors for traditional Ehrenfest are comparable to those for TAB and TAB-DMS, and smaller than those for TAB-w and TAB-w-DMS.  However, all TAB methods describe proper bifurcation of the trajectory swarm following coupling, whereas Ehrenfest is well known to continue to follow an unphysical mean-field PES following coupling.  These results strongly suggest that TAB is an appropriate method for modeling dynamics of polyatomic molecules in dense manifolds of intersecting states.  Therefore we are in the process of applying our \textit{ab initio} implementation of TAB to model long-lived coherences in highly-excited molecules.  Though agreement with exact quantum results is excellent, in general, we are also investigating the application of TAB using diabatic basis functions for collapsing, in order to assess the basis dependence of the method.

\section*{Acknowledgement(s)}

This work is dedicated to a wonderful friend and mentor, Piotr Piecuch, on the occasion of his 65th birthday. 

\section*{Funding}

The authors gratefully acknowledge support from the National Science Foundation under grant CHE-1954519 and the Institute for Advanced Computational Science (IACS).

\bibliographystyle{tfo}
\bibliography{citations}

\begin{thebibliography}{70}
\providecommand{\url}[1]{\texttt{#1}}
\providecommand{\urlprefix}{URL }

\bibitem{yarkony1996diabolical}
D. Yarkony,  REVIEWS OF MODERN PHYSICS  \textbf{68} (4), 985--1013 (1996).

\bibitem{levine2007isomerization}
B.G. Levine and T.J. Mart{\'\i}nez,  Annu. Rev. Phys. Chem.  \textbf{58}, 613--634 (2007).

\bibitem{zhu2016review}
X. Zhu and D.R. Yarkony,  MOLECULAR PHYSICS  \textbf{114} (13), 1983--2013 (2016).

\bibitem{schuurman2018dynamics}
M.S. Schuurman and A. Stolow,  Annual review of physical chemistry  \textbf{69}, 427--450 (2018).

\bibitem{levine2019conical}
B.G. Levine, M.P. Esch, B.S. Fales, D.T. Hardwick, W.T. Peng and Y. Shu,  Annual Review of Physical Chemistry  \textbf{70}, 21--43 (2019).

\bibitem{matsika2021chemrev}
S. Matsika,  CHEMICAL REVIEWS  \textbf{121} (15), 9407--9449 (2021).

\bibitem{yarkony2001dynamics}
D. Yarkony,  JOURNAL OF CHEMICAL PHYSICS  \textbf{114} (6), 2601--2613 (2001).

\bibitem{cederbaum2005shorttime}
L. Cederbaum, E. Gindensperger and I. Burghardt,  PHYSICAL REVIEW LETTERS  \textbf{94} (11) (2005).

\bibitem{tully1971trajectory}
J.C. Tully and R.K. Preston,  The Journal of chemical physics  \textbf{55} (2), 562--572 (1971).

\bibitem{tully1990molecular}
J.C. Tully,  The Journal of Chemical Physics  \textbf{93} (2), 1061--1071 (1990).

\bibitem{hammes1994proton}
S. Hammes-Schiffer and J.C. Tully,  The Journal of chemical physics  \textbf{101} (6), 4657--4667 (1994).

\bibitem{coker1995methods}
D.F. Coker and L. Xiao,  The Journal of chemical physics  \textbf{102} (1), 496--510 (1995).

\bibitem{tully1998mixed}
J. Tully,  Faraday Discussions  \textbf{110}, 407--419 (1998).

\bibitem{jasper2001photodissociation}
A.W. Jasper, M.D. Hack, A. Chakraborty, D.G. Truhlar and P. Piecuch,  The Journal of Chemical Physics  \textbf{115} (17), 7945--7952 (2001).

\bibitem{subotnik2016understanding}
J.E. Subotnik, A. Jain, B. Landry, A. Petit, W. Ouyang and N. Bellonzi,  Annual review of physical chemistry  \textbf{67}, 387--417 (2016).

\bibitem{wang2016recent}
L. Wang, A. Akimov and O.V. Prezhdo,  The journal of physical chemistry letters  \textbf{7} (11), 2100--2112 (2016).

\bibitem{ben2000ab}
M. Ben-Nun, J. Quenneville and T.J. Mart{\'\i}nez,  The Journal of Physical Chemistry A  \textbf{104} (22), 5161--5175 (2000).

\bibitem{worth2003full}
G.A. Worth and I. Burghardt,  Chemical physics letters  \textbf{368} (3-4), 502--508 (2003).

\bibitem{wu2003matching}
Y. Wu and V.S. Batista,  The Journal of chemical physics  \textbf{118} (15), 6720--6724 (2003).

\bibitem{curchod2018ab}
B.F. Curchod and T.J. Mart{\'\i}nez,  Chemical reviews  \textbf{118} (7), 3305--3336 (2018).

\bibitem{makri2015quantum}
N. Makri,  International Journal of Quantum Chemistry  \textbf{115} (18), 1209--1214 (2015).

\bibitem{walters2015quantum}
P.L. Walters and N. Makri,  The Journal of Physical Chemistry Letters  \textbf{6} (24), 4959--4965 (2015).

\bibitem{min2015coupled}
S.K. Min, F. Agostini and E.K. Gross,  Physical review letters  \textbf{115} (7), 073001 (2015).

\bibitem{min2017ab}
S.K. Min, F. Agostini, I. Tavernelli and E.K. Gross,  The journal of physical chemistry letters  \textbf{8} (13), 3048--3055 (2017).

\bibitem{curchod2018ct}
B.F. Curchod, F. Agostini and I. Tavernelli,  The European Physical Journal B  \textbf{91}, 1--12 (2018).

\bibitem{gossel2018coupled}
G.H. Gossel, F. Agostini and N.T. Maitra,  Journal of chemical theory and computation  \textbf{14} (9), 4513--4529 (2018).

\bibitem{vindel2022exact}
P. Vindel-Zandbergen, S. Matsika and N.T. Maitra,  The Journal of Physical Chemistry Letters  \textbf{13} (7), 1785--1790 (2022).

\bibitem{kouppel1984multimode}
H. K{\"o}uppel, W. Domcke and L.S. Cederbaum,  Advances in chemical physics  pp. 59--246 (1984).

\bibitem{guo2016accurate}
H. Guo and D.R. Yarkony,  Physical Chemistry Chemical Physics  \textbf{18} (38), 26335--26352 (2016).

\bibitem{miller1978classical}
W. Miller and C. McCurdy,  The Journal of Chemical Physics  \textbf{69} (11), 5163--5173 (1978).

\bibitem{meyera1979classical}
H.D. Meyera) and W.H. Miller,  The Journal of Chemical Physics  \textbf{70} (7), 3214--3223 (1979).

\bibitem{macias1982ab}
A. Mac{\'\i}as and A. Riera,  Physics Reports  \textbf{90} (5), 299--376 (1982).

\bibitem{micha1983self}
D.A. Micha,  The Journal of Chemical Physics  \textbf{78} (12), 7138--7145 (1983).

\bibitem{cotton2013symmetrical}
S.J. Cotton and W.H. Miller,  The Journal of Physical Chemistry A  \textbf{117} (32), 7190--7194 (2013).

\bibitem{isborn2007time}
C.M. Isborn, X. Li and J.C. Tully,  The Journal of chemical physics  \textbf{126} (13) (2007).

\bibitem{prezhdo1997mean}
O.V. Prezhdo and P.J. Rossky,  The Journal of chemical physics  \textbf{107} (3), 825--834 (1997).

\bibitem{prezhdo1997evaluation}
O.V. Prezhdo and P.J. Rossky,  The Journal of chemical physics  \textbf{107} (15), 5863--5878 (1997).

\bibitem{bedard2005mean}
M.J. Bedard-Hearn, R.E. Larsen and B.J. Schwartz,  The Journal of chemical physics  \textbf{123} (23) (2005).

\bibitem{subotnik2010augmented}
J.E. Subotnik,  The Journal of chemical physics  \textbf{132} (13) (2010).

\bibitem{hack2001natural}
M.D. Hack and D.G. Truhlar,  The Journal of Chemical Physics  \textbf{114} (21), 9305--9314 (2001).

\bibitem{zhu2004non}
C. Zhu, A.W. Jasper and D.G. Truhlar,  The Journal of chemical physics  \textbf{120} (12), 5543--5557 (2004).

\bibitem{zhu2004coherent}
C. Zhu, S. Nangia, A.W. Jasper and D.G. Truhlar,  The Journal of chemical physics  \textbf{121} (16), 7658--7670 (2004).

\bibitem{akimov2014coherence}
A.V. Akimov, R. Long and O.V. Prezhdo,  The Journal of chemical physics  \textbf{140} (19) (2014).

\bibitem{makhov2014ab}
D.V. Makhov, W.J. Glover, T.J. Martinez and D.V. Shalashilin,  The Journal of chemical physics  \textbf{141} (5) (2014).

\bibitem{freixas2018ab}
V.M. Freixas, S. Fernandez-Alberti, D.V. Makhov, S. Tretiak and D. Shalashilin,  Physical Chemistry Chemical Physics  \textbf{20} (26), 17762--17772 (2018).

\bibitem{fedorov2019nonadiabatic}
D.A. Fedorov and B.G. Levine,  The Journal of Physical Chemistry Letters  \textbf{10} (16), 4542--4548 (2019).

\bibitem{kaufman2023coherence}
B. Kaufman, P. Marquetand, T. Rozgonyi and T. Weinacht,  PHYSICAL REVIEW LETTERS  \textbf{131} (26) (2023).

\bibitem{holland2024auger}
D.M.P. Holland, J. Suchan, J. Janos, C. Bacellar, L. Leroy, T.R. Barillot, L. Longetti, M. Coreno, M. de~Simone, C. Grazioli, M. Chergui, E. Muchova and R.A. Ingle,  PHYSICAL CHEMISTRY CHEMICAL PHYSICS  \textbf{26} (21) (2024).

\bibitem{alzubeidi2023solvatede}
A. Al-Zubeidi, B. Ostovar, C.C. Carlin, B.C. Li, S.A. Lee, W.Y. Chiang, N. Gross, S. Dutta, A. Misiura, E.K. Searles, A. Chakraborty, S.T. Roberts, J.A. Dionne, P.J. Rossky, C.F. Landes and S. Link,  PROCEEDINGS OF THE NATIONAL ACADEMY OF SCIENCES OF THE UNITED STATES OF AMERICA  \textbf{120} (3) (2023).

\bibitem{hetherington2023cooling}
C.V. Hetherington, T.M.N. Mohan, R.W. Tilluck, W.F. Beck and B.G. Levine,  JOURNAL OF PHYSICAL CHEMISTRY LETTERS  \textbf{14} (51), 11651--11658 (2023).

\bibitem{craig2005trajectory}
C.F. Craig, W.R. Duncan and O.V. Prezhdo,  Physical review letters  \textbf{95} (16), 163001 (2005).

\bibitem{dou2015surface}
W. Dou, A. Nitzan and J.E. Subotnik,  The Journal of chemical physics  \textbf{142} (23) (2015).

\bibitem{dou2015frictional}
W. Dou, A. Nitzan and J.E. Subotnik,  The Journal of chemical physics  \textbf{143} (5) (2015).

\bibitem{ouyang2015surface}
W. Ouyang, W. Dou and J.E. Subotnik,  The Journal of Chemical Physics  \textbf{142} (8) (2015).

\bibitem{esch2020decoherence}
M.P. Esch and B.G. Levine,  The Journal of Chemical Physics  \textbf{153} (11) (2020).

\bibitem{esch2020state}
M.P. Esch and B.G. Levine,  The Journal of Chemical Physics  \textbf{152} (23) (2020).

\bibitem{suchan2024predict}
J. Suchan, F. Liang, A.S. Durden and B.G. Levine,  JOURNAL OF CHEMICAL PHYSICS  \textbf{160} (13) (2024).

\bibitem{esch2021accurate}
M.P. Esch and B.G. Levine,  The Journal of Chemical Physics  \textbf{155} (21) (2021).

\bibitem{swope1982computer}
W.C. Swope, H.C. Andersen, P.H. Berens and K.R. Wilson,  The Journal of chemical physics  \textbf{76} (1), 637--649 (1982).

\bibitem{blanes2006symplectic}
S. Blanes, F. Casas and A. Murua,  The Journal of chemical physics  \textbf{124} (23) (2006).

\bibitem{bittner1995decoherence}
E. BITTNER and P. ROSSKY,  JOURNAL OF CHEMICAL PHYSICS  \textbf{103} (18), 8130--8143 (1995).

\bibitem{subotnik2011rescaling}
J.E. Subotnik and N. Shenvi,  JOURNAL OF CHEMICAL PHYSICS  \textbf{134} (2) (2011).

\bibitem{martens2019rescaling}
C.C. Martens,  JOURNAL OF PHYSICAL CHEMISTRY A  \textbf{123} (5), 1110--1128 (2019).

\bibitem{plasser2019rescaling}
F. Plasser, S. Mai, M. Fumanal, E. Gindensperger, C. Daniel and L. Gonzalez,  JOURNAL OF CHEMICAL THEORY AND COMPUTATION  \textbf{15} (9), 5031--5045 (2019).

\bibitem{barbatti2021rescaling}
M. Barbatti,  JOURNAL OF CHEMICAL THEORY AND COMPUTATION  \textbf{17} (5), 3010--3018 (2021).

\bibitem{nangia2005balanced}
S. Nangia, D. Truhlar, M. McGuire and P. Piecuch,  JOURNAL OF PHYSICAL CHEMISTRY A  \textbf{109} (51), 11643--11646 (2005).

\bibitem{levine2006tddft}
B. Levine, C. Ko, J. Quenneville and T. Martínez,  MOLECULAR PHYSICS  \textbf{104} (5-7), 1039--1051 (2006).

\bibitem{granovsky2011xms}
A.A. Granovsky,  JOURNAL OF CHEMICAL PHYSICS  \textbf{134} (21) (2011).

\bibitem{gozem2014shape}
S. Gozem, F. Melaccio, A. Valentini, M. Filatov, M. Huix-Rotllant, N. Ferre, L. Manuel~Frutos, C. Angeli, A.I. Krylov, A.A. Granovsky, R. Lindh and M. Olivucci,  JOURNAL OF CHEMICAL THEORY AND COMPUTATION  \textbf{10} (8), 3074--3084 (2014).

\bibitem{park2021butadiene}
W. Park, J. Shen, S. Lee, P. Piecuch, M. Filatov and C.H. Choi,  JOURNAL OF PHYSICAL CHEMISTRY LETTERS  \textbf{12} (39), 9720--9729 (2021).

\end{thebibliography}

\end{document}